\documentclass[showpacs,preprintnumbers,amsmath,amssymb]{revtex4}
\usepackage{amsmath,amssymb,graphics,epsfig,subfigure}
\usepackage{color}

\begin{document}

\title{Quasinormal Modes and Van der Waals like phase transition of
charged AdS black holes in Lorentz symmetry breaking massive gravity}

\author{Bin Liang \footnote{liangb2016@lzu.edu.cn},
        Shao-Wen Wei \footnote{ weishw@lzu.edu.cn, corresponding author},
        Yu-Xiao Liu \footnote{liuyx@lzu.edu.cn, corresponding author}
}
\affiliation{Institute of Theoretical Physics, Lanzhou University, Lanzhou 730000, China\\
             Center for Gravitational Physics, Lanzhou University, Lanzhou 730000, China}

\begin{abstract}
Using the quasinormal modes of a massless scalar perturbation, we investigate the small/large black hole phase transition in the Lorentz symmetry breaking massive gravity. We mainly focus on two issues: i) the sign change of slope of the quasinormal mode frequencies in the complex-$\omega$ diagram; ii) the behaviors of the imaginary part of the quasinormal mode frequencies along the isobaric or isothermal processes. For the first issue, our result shows that, at low fixed temperature or pressure, the phase transition can be probed by the sign change of slope. While increasing the temperature or pressure to some certain values near the critical point, there will appear the deflection point, which indicates that such method may not be appropriate to test the phase transition. In particular, the behavior of the quasinormal mode frequencies for the small and large black holes tend to the same at the critical point. For the second issue, it is shown that the non-monotonic behavior is observed only when the small/large black hole phase transition occurs. Therefore, this property can provide us with an additional method to probe the phase transition through the quasinormal modes.
\end{abstract}



\maketitle

\section{Introduction}
\label{1}

Black hole thermodynamics has always been a hot topic due to the pioneer work by Hawking and Page \cite{Hawking}, which demonstrated a thermodynamic phase transition between a large Schwarzschild-AdS black hole and a thermal AdS space. Later, the first-order phase transition between small and large black holes in the canonical ensemble was discovered in rotating and charged AdS black holes \cite{Chamblin,Chamblin2,Caldarelli}.

Considering the AdS/CFT correspondence, recent study of black hole thermodynamics has been generalized to extended phase space \cite{Dolan,Dolan2}, which identifies the cosmological constant as the thermodynamic pressure. In this case, the mass of an AdS black hole is referred to as the enthalpy rather than the internal energy \cite{Kastor}. Moreover, in Ref. \cite{Kubiznak}, it was found that there exists a small/large black hole thermodynamic phase transition in four-dimensional Reissner-Nordstrom (RN) AdS black holes, which is extremely analogous to the liquid-gas phase transition in the van der Waals (vdW) fluid in the canonical ensemble. It was also found that they share the same critical exponents. Subsequently, more phase structures of AdS black holes were disclosed, such as reentrant phase transition and triple point \cite{Gunasekaran,Hendi,Altamirano,Altamirano2,Zhao,Zou,Cai,Wei,Mo,Hennigar,Wei2,Zeng,Frassino,
Xu,Xu2,Hendi2,Cheng,Wei3,Zeng2,MajhiSamanta,HendiPanah}.

On the other hand, the quasinormal modes (QNMs) are corresponded to the dynamical perturbations in the surrounding geometry of a black hole \cite{Nollert,Kokkotas,Konoplya}. Since the QNMs depend on the hairy black hole parameters, it is natural to use it to probe the thermodynamic phase transition of the black hole. In Ref. \cite{Liu}, Liu, Zou, and Wang found that there exists a dramatic change in the slope of the QNM frequencies in small and large black holes around the point where the first-order phase transition takes place. The study was also extended to other AdS black holes \cite{Mahapatra,Chabab,Zou2,Prasia,Shi2}. These results imply that the QNMs can be used as a dynamic probe for the first-order black hole phase transition. More works can be found in Refs. \cite{Koutsoumbas,Shen,Rao,Berti,He,Miranda,Cai2,Zou3,Konoplya:2002ky,Konoplya:2002zu,
CardosoLemos,CardosoLemos2,Chakraborty}. It's worthwhile noting that there is a correspondence between the QNMs and the decay of the perturbations in the dual conformal field theory. The first work was carried out in Ref .\cite{Horowitz} through numerical computation. It was found that the timescale for the approach to the thermal equilibrium in the CFT is simply the reciprocal of absolute value of the imaginary part of the lowest QNM frequency and the imaginary part has good linearity on the scale of the temperature.

In this paper, we would like to examine the QNMs and phase transition of charged AdS black holes in Lorentz symmetry breaking (LSB) massive gravity, which has some novel and interesting results. Its symmetry is broken by a space-time dependent condensate of scalar fields. And interestingly, there exist two propagating degrees of freedom for the scalar sector. Although there are some ghost-like instabilities in the maximally symmetric spaces, they can be avoided in FRW background. At low fixed temperature or pressure, we find that the slope of the QNM frequencies will change at the phase transition point. However, when approaching to the critical point, this property will not hold anymore. Fortunately, we find that there exists non-monotonic behavior in the $\omega_i-T$ (or $\omega_i-P$) diagram. Therefore, the non-monotonic behaviors provide us with an additional probe of the phase transition.

This paper is organized as follows. In Sec. \ref{2}, we will briefly review phase transition of the charged AdS black holes in the LSB massive gravity. In Sec. \ref{3}, we numerically calculate the QNM frequencies for the black holes around the phase transition points along the isobaric and isothermal processes, respectively. The study of the non-monotonic behavior of the imaginary part of the QNM frequencies is given in Sec. \ref{4}. The final section is devoted to the conclusions and discussions.

\section{Phase transition and thermodynamics}
\label{2}

Massive gravity theory was firstly developed by Fierz and Pauli in 1939 \cite{Fierz}. Here we consider the LSB massive gravity theory. For the details of this massive gravity theory, we refer readers to \cite{Dubovsky,Rubakov}.

The LSB massive gravity was systematically studied in Ref. \cite{Blas}, and its action reads
\begin{eqnarray}
{S}=\int{d^4x\sqrt{-g}\left[-{M_{pl}}^2{\cal R}+\Omega^4{\cal F}(X,W^{ij})\right]}, \label{action}
\end{eqnarray}
where $M_{pl}$ is the Plank mass, $\cal R$ is the scalar curvature of the space-time geometry,
$\cal F$ is the function of $X$ and $W^{ij}$:
\begin{eqnarray}
{X}&=&\frac{\partial^{\mu}\Phi^0\partial_{\mu}\Phi^0}{\Omega^4},\\ \label{para1}
{W^{ij}}&=&\frac{\partial^{\mu}\Phi^{i}\partial_{\mu}\Phi^{j}}{\Omega^4}-
\frac{\partial^{\mu}\Phi^{i}\partial_{\mu}\Phi^0\partial^{\nu}\Phi^{j}\partial_{\nu}\Phi^0}{\Omega^4}. \label{para2}
\end{eqnarray}
Here, the parameter $\Omega$ has the dimension of mass, and the scalar fields $\Phi^0, \Phi^{i}$ are Goldstone fields which break the Lorentz symmetry spontaneously so that the graviton gets a mass. As discussed in \cite{Bebronne}, analytical solution is impossible to find for a generic function $\cal F$. However, the static spherically symmetric metric solution of a charged AdS black hole in this LSB massive gravity can be obtained for a specific function $\cal F$ introduced in \cite{Bebronne}.
\begin{eqnarray}
ds^2=-f(r)dt^2+\frac{1}{f(r)}dr^2+r^2(d\theta^2+\sin^2\theta d\varphi^2), \label{metric}
\end{eqnarray}
where the metric function reads
\begin{eqnarray}
f(r)=1-\frac{2M}{r}-\gamma\frac{Q^2}{r^{\lambda}}-\frac{\Lambda r^2}{3}. \label{solution}
\end{eqnarray}
When $\lambda<1$, the ADM mass will become divergent, which is not allowed \cite{Fernando}. So we choose $\lambda>1$. Importantly, when $\gamma=1$, the metric solution is similar to the Schwarzschild AdS black hole with a single horizon where the phase transition is trivial. When $\gamma=-1$, the metric solution is much like RN-AdS black hole which admits a nontrivial small/large black hole phase transition \cite{Fernando}.
Since we are interested in the nontrivial phase transition, we choose $\gamma=-1$ in the following discussion. In addition, we use the geometric units $G=\hbar=c=k_B=1$ for simplicity. For the asymptotically flat case, the strong gravitational lensing was studied in Ref. \cite{ZhangChen}.

The temperature of this black hole can be calculated as
\begin{eqnarray}
T=\frac{f'(r_h)}{4\pi}=\frac{1}{4\pi}\left(\frac{2M}{r_h^2}+\frac{\gamma Q^2\lambda}{r_h^{\lambda+1}}-\frac{2\Lambda r_h}{3}\right), \label{T}
\end{eqnarray}
where $r_h$ is the radius of the outer event horizon and $M$ is the mass of the black hole. Using $f(r_h)=0$, the mass can be obtained
\begin{eqnarray}
M=\frac{r_h}{2}-\frac{\gamma Q^2}{2r_h^{\lambda-1}}-\frac{\Lambda r_h^3}{6}. \label{M}
\end{eqnarray}
In the extended phase space, the cosmological constant corresponds to thermodynamic pressure,
\begin{eqnarray}
\Lambda=-8\pi P. \label{P}
\end{eqnarray}
To compare with the vdW fluid equation, we identify the specific volume $v$ of the black hole fluid with the horizon radius as \cite{Kubiznak}
\begin{eqnarray}
v=2r_h.\label{v}
\end{eqnarray}
With the substitution of Eqs.~(\ref{T})-(\ref{v}), the equation of state of the black hole can be obtained
\begin{eqnarray}
P=-\frac{1}{2\pi v^2}+\frac{T}{v}+\frac{Q^2(\lambda-1)2^{\lambda}}{2\pi v^{2+\lambda}}. \label{eos}
\end{eqnarray}
The critical point can be determined by
\begin{eqnarray}
\frac{\partial P}{\partial v}\Big|_{T=T_c, v=v_c}
=\frac{\partial^2 P}{\partial v^2}\Big|_{T=T_c, v=v_c}=0, \label{eq:5s}
\end{eqnarray}
which leads to \cite{Fernando}
\begin{eqnarray}
v_c&=& \big[Q^2(\lambda^2-1)(\lambda+2)2^{\lambda-1}\big]^{\frac{1}{\lambda}},\nonumber\\
T_c&=&\frac{\lambda}{\pi(\lambda+1)}\big[Q^2(\lambda^2-1)(\lambda+2)2^{\lambda-1}\big]^{\frac{-1}{\lambda}},\nonumber\\
P_c&=&\frac{\lambda}{2\pi(\lambda+2)}\big[Q^2(\lambda^2-1)(\lambda+2)2^{\lambda-1}\big]^{\frac{-2}{\lambda}}.\label{cri}
\end{eqnarray}
Noting that when $\lambda=2$, the critical point given above is exactly
the same as that of the RN-AdS black hole \cite{Kubiznak}. It is clear that the graviton mass significantly modifies this behavior, and a non-zero graviton mass admits the possibility of critical behavior for $\lambda\neq2$. In Table~\ref{tab}, we can see effects of different values of $\lambda$ on critical pressure and temperature around $\lambda=2$  when the black hole charge $Q$=1.

\begin{table}[h]
\begin{center}
\begin{tabular}{c|c|c|c|c|c}
  \hline\hline
  $\lambda$ & 1.8 & 1.9 & 2.0 & 2.1 & 2.2 \\\hline
  $P_c$ $(10^{-3})$ & 3.77 & 3.50 & 3.32 & 3.20 & 3.12 \\\hline
  $T_c$ $(10^{-2})$ & 4.58 & 4.43 & 4.33 & 4.27 & 4.24 \\\hline\hline
\end{tabular}
\caption{Different values of $P_c$ and $T_c$ with different $\lambda$. The black hole charge is set to $Q=1$.}\label{tab}
\end{center}
\end{table}

Global stability of a black hole can be determined by studying the corresponding Gibbs free energy. In the extended phase space, $M$ is interpreted as the enthalpy rather than the internal energy. So with the substitution of Eq.~(\ref{M}), the corresponding Gibbs free energy is
\begin{equation}
G(Q,P)=M-TS=\frac{r_h}{4}-\frac{\gamma Q^2(1+\lambda)}{4r_h^{\lambda-1}}-\frac{2\pi P r_h^3}{3}. \label{G}
\end{equation}
With the help of Eq.~(\ref{eos}), the Gibbs free energy can be expressed in terms of $r_h$, $T$, and $Q$
\begin{equation}
G(Q,T)=M-TS=\frac{1}{6}\left[2r_h-2\pi r_h^2 T-Q^2 r_h^{1-\lambda}\gamma(2+\lambda)\right]. \label{G2}
\end{equation}
It is shown in Ref. \cite{Fernando} that this AdS black hole admits a phase transition of the vdW type. Below the critical point (\ref{cri}), the Gibbs free energy (\ref{G2}) shows a swallow tail behavior indicating the existence of a first-order phase transition. And with the increase of the temperature or pressure, the swallow tail behavior shrinks and exactly disappears at the critical point of second-order, and completely disappears beyond the critical point.

\section{Quasinormal mode frequencies}
\label{3}

In this section, we would like to investigate the QNMs and phase transition for the massive black holes. 

In the asymptotically flat spacetime, where the cosmological constant $\Lambda$ vanishes, the QNMs of scalar perturbations and Dirac field perturbations have been studied in Refs. \cite{Fernando2,Fernando3}. For different values of $\lambda$, the QNMs were calculated by using the sixth order WKB approach. Moreover, adopting the Poschl-Teller potential approximation, the QNMs can be obtained analytically. Based on this idea, the analytical QNMs were obtained for the massive black holes \cite{Fernando2,Fernando3}.

Here we aim to calculate the QNMs of massless scalar field perturbation for the massive black hole in asymptotically AdS spacetime. For this non-vanishing cosmological constant case, there is no the analytical result, thus we will use the shooting method to obtain the QNMs. Then the relation between the QNMs and the thermodynamic phase transition will be examined.

Here we just take the radial part of the perturbation, $\Phi(r,t)=\phi(r)e^{-i\omega t}$, which is governed by the Klein-Gordon equation,
\begin{eqnarray}
\frac{1}{\sqrt{-g}}\partial_\mu\left(\sqrt{-g}g^{\mu\nu}\partial_\nu\Phi(r,t)\right)=0. \label{KG}
\end{eqnarray}
Substituting the metric (\ref{metric}) and the decomposition of the perturbation $\Phi(r,t)$ into Eq.~(\ref{KG}), one gets
\begin{eqnarray}
\phi''(r)+\left(\frac{f'(r)}{f(r)}+\frac{2}{r}\right)\phi'(r)+\frac{\omega^2\phi(r)}{f(r)^2}=0, \label{KG1}
\end{eqnarray}
where the QNM frequency  $\omega=\omega_r + i\omega_{i}$, and $\omega_r$ and $\omega_{i}$ correspond to oscillation frequency and damping time, respectively. For this black hole system, we need to impose boundary conditions both at the horizon and at the space infinity. \\
Then we can define $\phi(r)$ as $\varphi(r)\exp[-i\int\frac{\omega}{f(r)}dr]$, where $\exp[-i\int\frac{\omega}{f(r)}dr]$ asymptotically approaches to ingoing wave
near the horizon. Thus, Eq.~(\ref{KG1}) takes the form
\begin{equation}
\varphi''(r)+\varphi'(r)\left[\frac{f'(r)}{f(r)}-\frac{2i\omega}{f(r)}+\frac{2}{r}\right]
-\frac{2i\omega}{r f(r)}\varphi(r)=0. \label{omegaKG}
\end{equation}
Without loss of generality, we can set $\varphi(r)=1$ when $r\rightarrow r_h$. Thanks to the infinite effective potential at the AdS boundary when $r\rightarrow\infty$, we set $\varphi(r\rightarrow\infty)=0$. With these boundary conditions,
we can numerically solve Eq.~(\ref{omegaKG}) and find the QNM frequencies by using the shooting method.

We are going to study whether the signature of phase transition in charged AdS black holes can be reflected in the dynamical QNMs behavior in the massless scalar perturbation. We will examine the dynamical perturbations in two processes, namely the isobaric process and the isothermal process. In our following numerical calculations, we set $Q=1$ and $\gamma=-1$.

\subsection{Isobaric phase transition}

\begin{figure}
\subfigure[]{\label{1a}
\includegraphics[width=6cm]{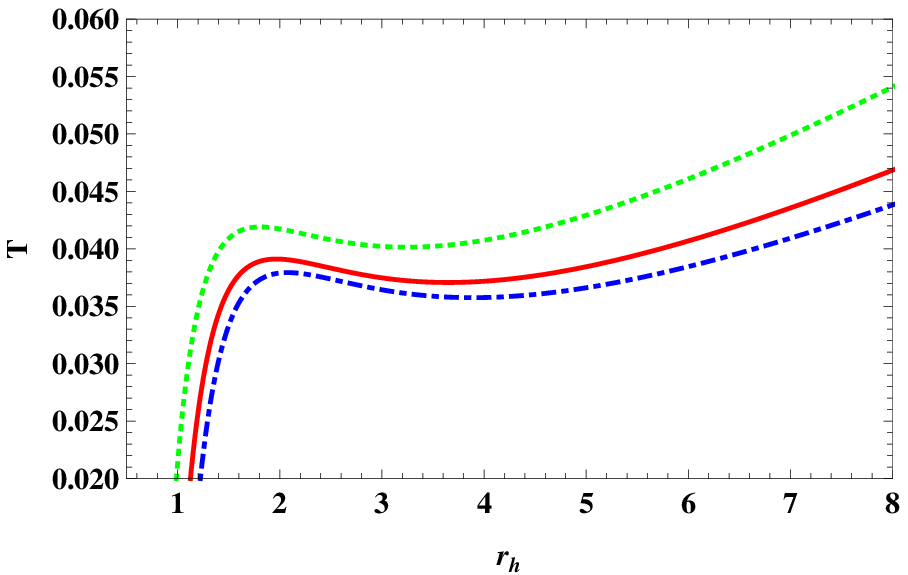}}
\subfigure[]{\label{1b}
\includegraphics[width=6cm]{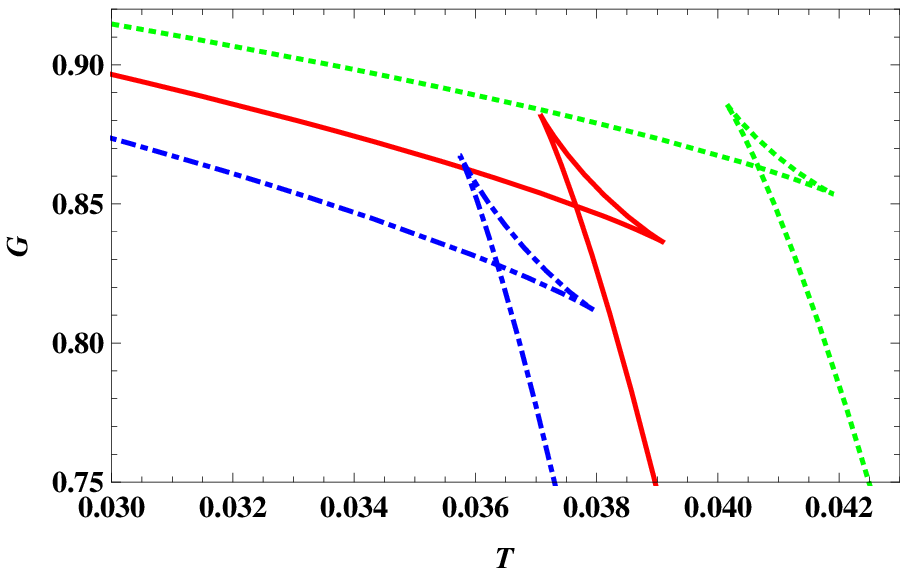}}
\caption{(a) $T$-$r_h$ diagram and (b) $G-T$ diagram for $P<P_c$. The three lines correspond to $\lambda=1.8$ (green dotted line), $\lambda=2.0$ (red solid line), and $\lambda=2.2$ (blue dot dashed line).}
\label{fig1}
\end{figure}

In this case, we choose a fixed pressure to study the thermodynamic phase transition. We plot the temperature $T$ in Fig.~\ref{1a} for $P<P_{c}$. And the result shows that there exist the non-monotonic behaviors, which indicate the 
existence of the first-order small/large black hole phase transition. The $G-T$ diagram is also presented in Fig.~\ref{1b}. Obviously, there display the swallow tail behaviors, which also imply the first-order small/large black hole phase transition. It is noticed that the intersection point denotes the coexistence phase of the small and large black holes. The critical pressure $P_c$ can be achieved by solving $\partial_{r_h}T=\partial_{r_h, r_h}T=0$.

\begin{figure}
\subfigure[]{\label{2a}
\includegraphics[width=6cm]{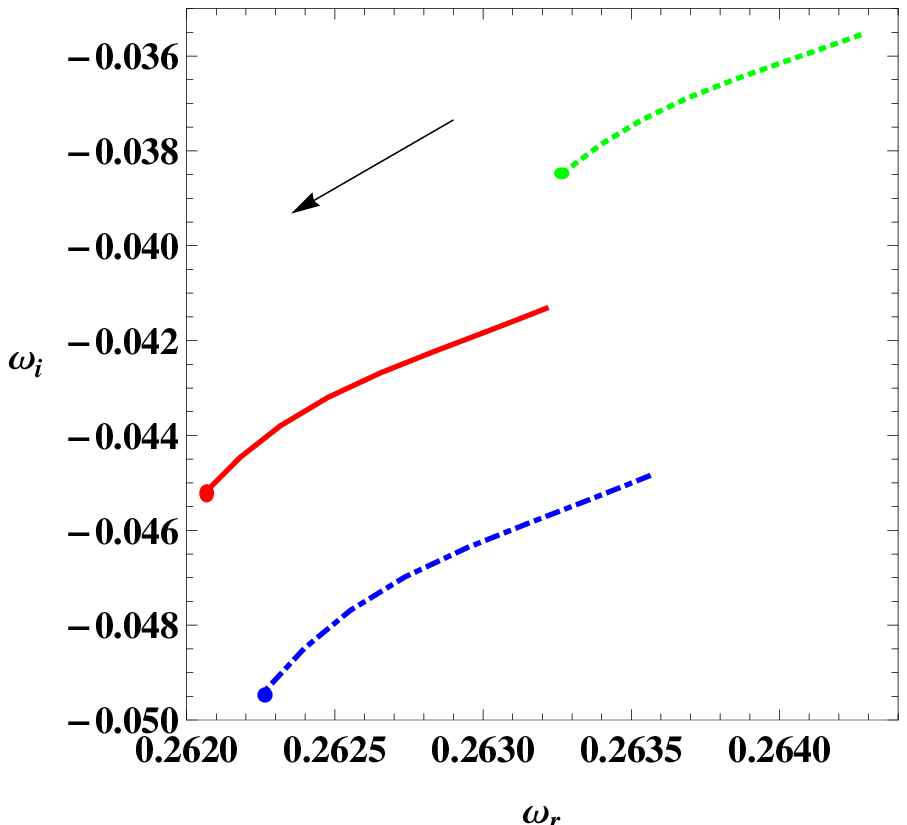}}
\subfigure[]{\label{2b}
\includegraphics[width=6cm]{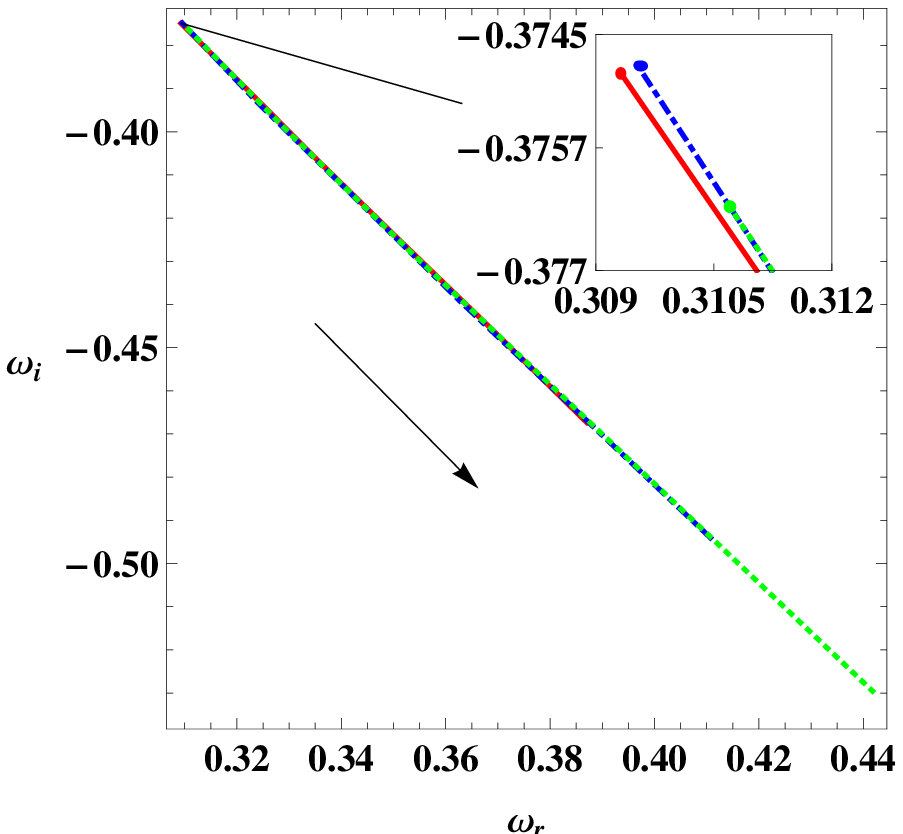}}
\caption{The behavior of the QNM frequencies in the complex-$\omega$ diagram with $P=0.0015$. (a) Small black holes. (b) Large black holes. The three lines correspond to $\lambda=1.8$ (green dotted line), $\lambda=2.0$ (red solid line), and $\lambda=2.2$ (blue dot dashed line). Solid points mean phase transition points. The arrows indicate the increase of the black hole horizon. The critical pressure $P_{c}$=0.00377, 0.00332, and 0.00312 for $\lambda$=1.8, 2.0, and 2.2, respectively. Thus the pressure $P=0.0015$ we choose is below the critical value.}\label{fig2}
\end{figure}

In Table~\ref{tab1}, we further list the frequencies of the massless scalar perturbation around the small and large black holes near the first-order phase transition. We choose the fixed pressure $P=0.0015$ for different values of $\lambda$=1.8, 2.0, 2.2. From Table I, we can find that the pressure we choose is below the critical values. So there exists the first order small-large black hole phase transition. It is noticed that, when $\lambda=2.0$, the situation is exactly the same as that of the RN-AdS black hole. While when $\lambda \neq 2.0$, it will deviate from the RN-AdS black hole case. For the small black hole phase, when the horizon radius gets smaller and smaller, the temperature decreases. The absolute value of the imaginary part of the QNM frequency decreases while the real part one changes very little. For the large black hole phase, the radius increases with the temperature. And the perturbation becomes more oscillatory due to the bounce effect of the AdS boundary and more damped due to the absorbtion of larger black hole \cite{Liu}. The behavior of the QNMs for the small and large  black holes with fixed pressure $P$=0.0015 can be found in Fig.~\ref{fig2}, and the arrow indicates the increase of the radius. We can clearly see that for the small black hole phase, it has a positive slope. While for the large black hole phase, the slope becomes negative. This result implies that we can use the sign change of the slope as a probe for the thermodynamic phase transition in the LSB massive gravity.

\begin{figure}
\subfigure[]{\label{3a}
\includegraphics[width=6cm]{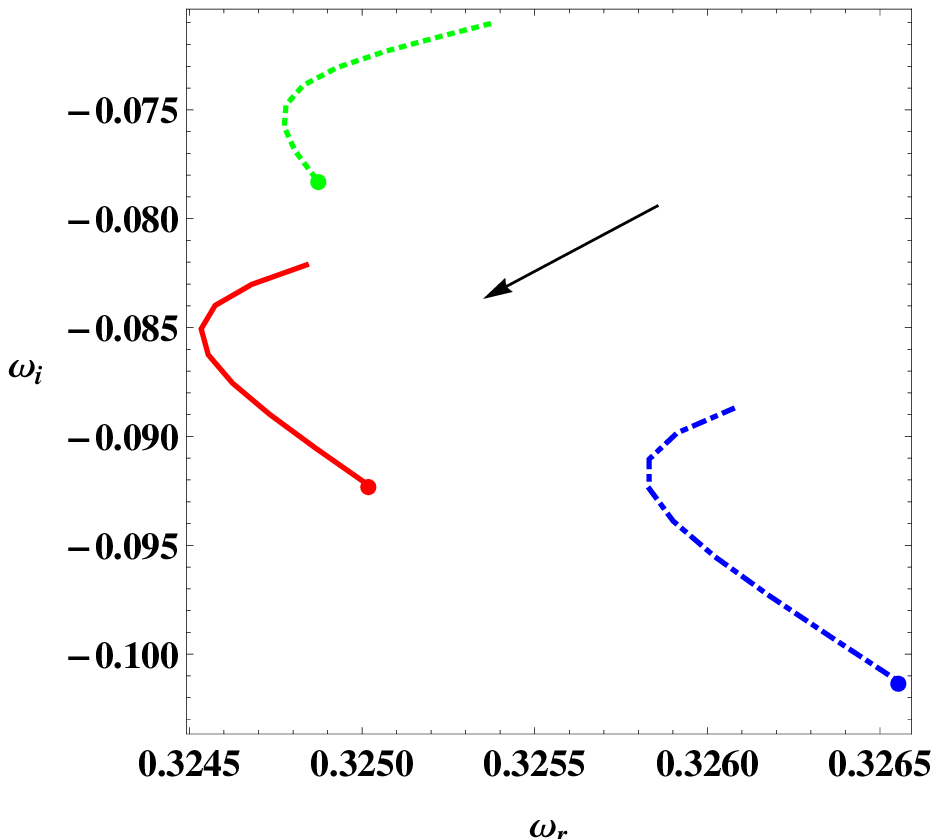}}
\subfigure[]{\label{3b}
\includegraphics[width=6cm]{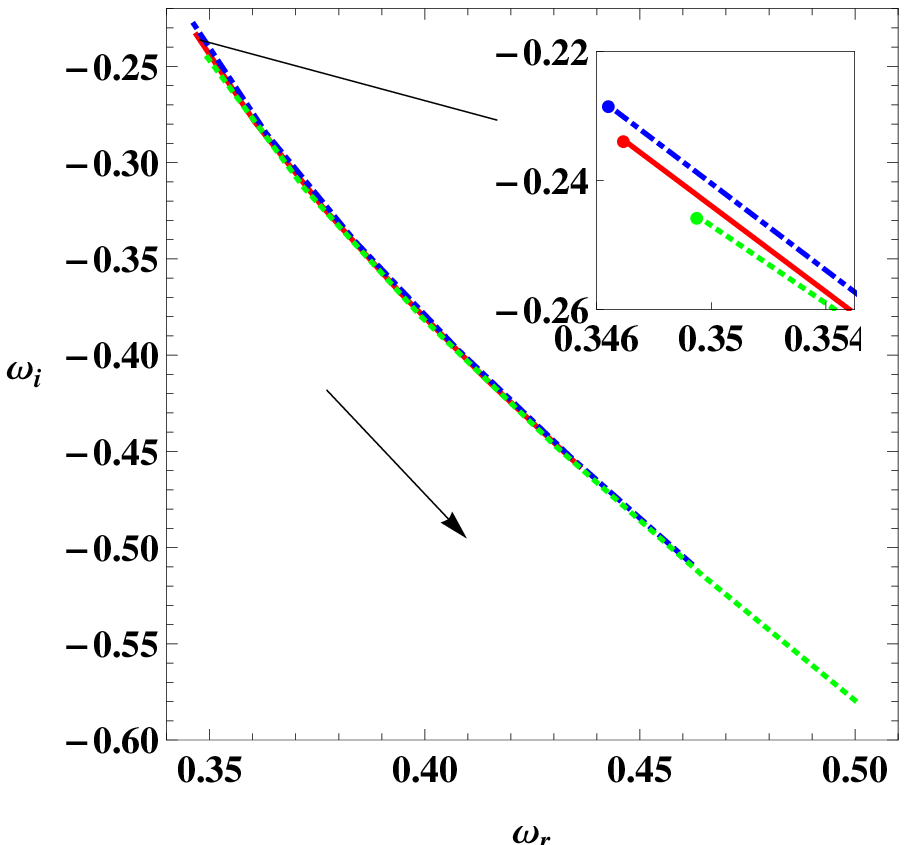}}
\caption{The behavior of the QNM frequencies in the complex-$\omega$ diagram with $P=0.0025$, which is still below the critical values of $\lambda$=1.8, 2.0, and 2.2. (a) Small black holes. (b) Large black holes. The three lines correspond to $\lambda=1.8$ (green dotted line), $\lambda=2.0$ (red solid line) and $\lambda=2.2$ (blue dot dashed line). Solid point means phase transition point. The arrows indicate the increase of the black hole horizon.}\label{fig3}
\end{figure}

Next, we increase the fixed pressure to $P=0.0025$, while which is still below the critical values. Then the strange behavior of the QNM frequencies appears. The QNM frequencies are given in Fig.~\ref{fig3}. From it, one can find the inflection point in the complex-$\omega$ diagram in Fig.~\ref{3a}. Thus the QNM frequencies for the small black hole phase do not have definite slope. So it seems that this method is not appropriate to probe the black hole phase transition for high pressure near the critical pressure. In fact, this result was also observed in Ref. \cite{Mahapatra}.

\begin{figure}
\includegraphics[width=7cm]{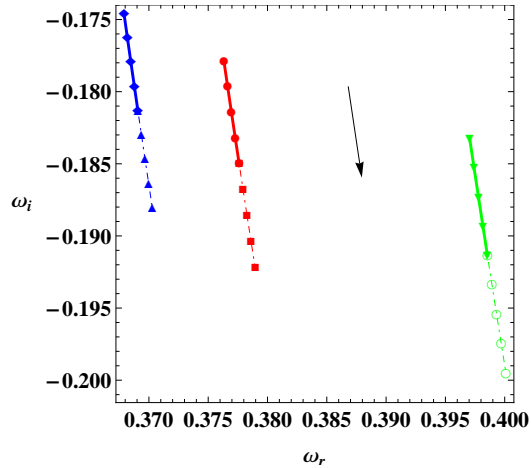}
\caption{The behavior of the QNM frequencies for large (dashed) and small (solid)
black holes in the complex-$\omega$ diagram with $P=P_c$. The three lines correspond to $\lambda=1.8$ (green), $\lambda=2.0$ (red) and $\lambda=2.2$ (blue). The arrow indicates the increase of the black hole horizon. The critical pressure $P_{c}$=0.00377, 0.00332, and 0.00312 for $\lambda$=1.8, 2.0, and 2.2, respectively.}\label{fig4}
\end{figure}

Moreover, at the critical pressure $P=P_c$, a second-order phase transition occurs. For different values of $\lambda$=1.8, 2.0, and 2.2, the QNM frequencies of the small and large black hole phases are plotted in Fig.~\ref{fig4}.
We can see that the QNM frequencies of the two black hole phases possess the same behavior with the increase of black hole horizon. Therefore, this slope method is also problematic at the critical point.

In a word, this method of sign change of slope is effective for probing the thermodynamic phase transition only at low fixed pressure.

\subsection{Isothermal phase transition}

\begin{figure}
\subfigure[]{\label{5a}
\includegraphics[width=6cm]{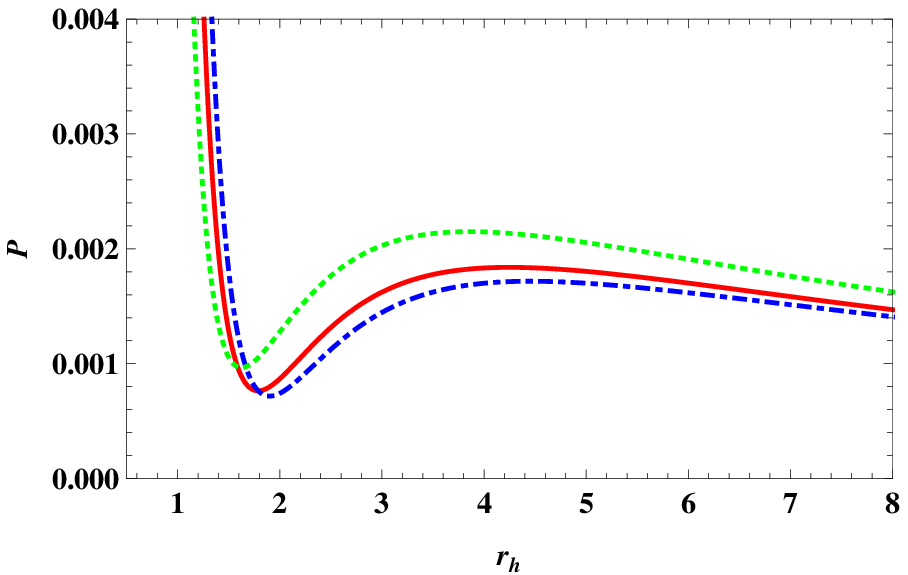}}
\subfigure[]{\label{5b}
\includegraphics[width=6cm]{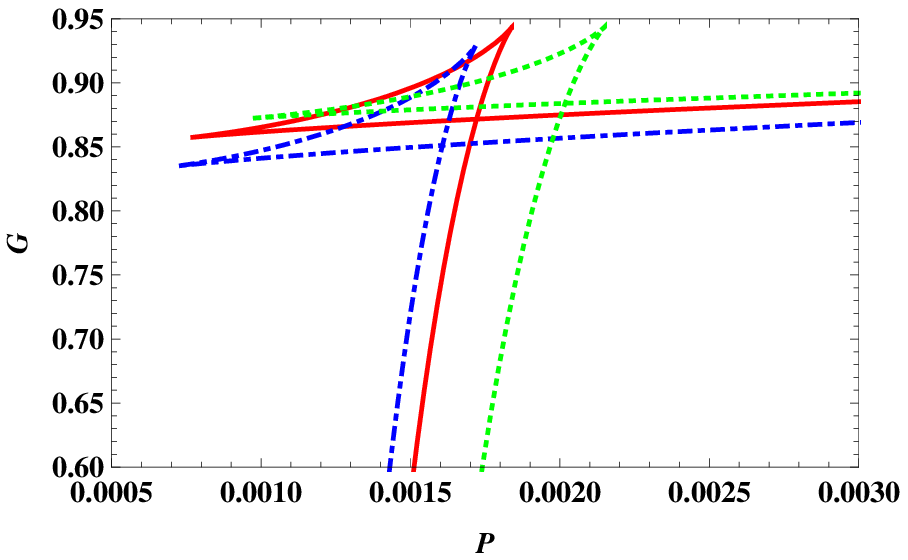}}
\caption{ (a)$P-r_h$ diagrams and (b) $G-P$ diagram for $T<T_c$. The three lines correspond to $\lambda=1.8$ (green dotted line), $\lambda=2.0$ (red solid line), and $\lambda=2.2$ (blue dot dashed line).}\label{fig5}
\end{figure}

For the isothermal process, we adopt the similar analysis as that of the above isobaric process. For $T<T_{c}$, we plot the behaviors of the pressure $P$ and the Gibbs free energy in Fig.~\ref{fig5}. Obviously, both the non-monotonic behavior of the pressure and the swallow tail behavior of the Gibbs free energy indicate that there exists a first-order black hole phase transition for $T<T_{c}$ in the LSB massive gravity.

In Table~\ref{tab2}, we further list the frequencies of the massless scalar perturbation near the small and large black hole phases for the first-order phase transition with fixed temperature $T=0.03$, which smaller than the critical values of the temperature with $\lambda$=1.8, 2.0, and 2.2. For the small black hole phase, when decreasing the radius of the black hole horizon, we find the QNM frequencies show more oscillation and more decay, which is very different from the isobaric process. The changes of the pressure $P$ and radius affect the frequencies in the opposite way, but in the end, we find the effect of the pressure $P$ overwhelms that of the horizon radius $r_h$, which leads to more oscillation due to the AdS boundary. For the large black hole phase, the radius decreases with the pressure. The real part slightly increases while the absolute value of the imaginary part decreases with the pressure, which is caused again by the absorbtion of the smaller black hole.

\begin{figure}
\subfigure[]{\label{6a}
\includegraphics[width=6cm]{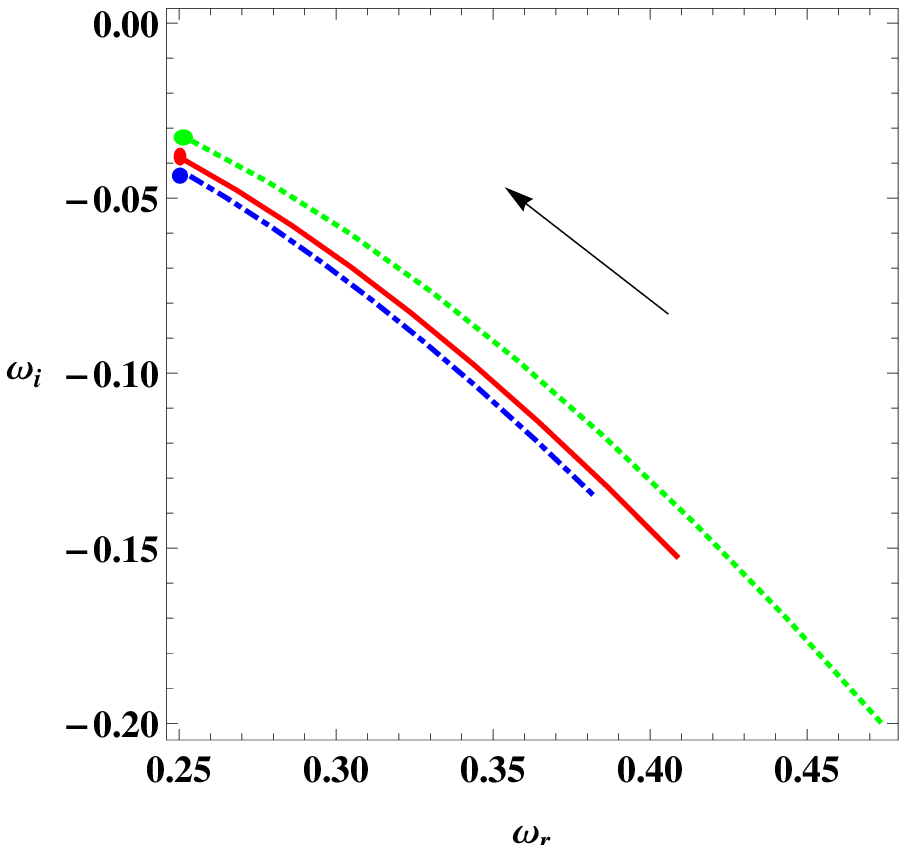}}
\subfigure[]{\label{6b}
\includegraphics[width=6cm]{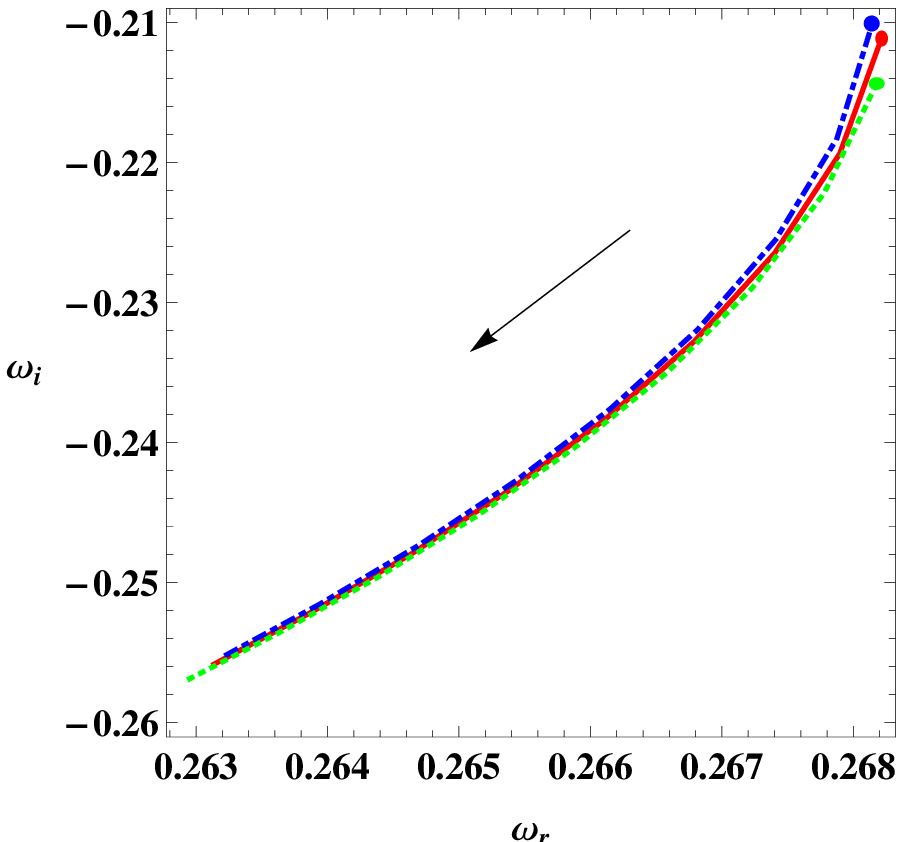}}
\caption{The behavior of the QNM frequencies in the complex-$\omega$ diagram with $T=0.03$. (a) Small black holes. (b) Large black holes. The three lines correspond to $\lambda=1.8$ (green dotted line), $\lambda=2.0$ (red solid line) and $\lambda=2.2$(blue dot dashed line). Solid point means phase transition point. The arrows indicate the increase of the black hole horizon. The critical temperature $T_{c}$=0.0458, 0.0433, and 0.0424 for $\lambda$=1.8, 2.0, and 2.2, respectively. Therefore, $T=0.03$ is below the critical temperature.}\label{fig6}
\end{figure}

\begin{figure}
\subfigure[]{\label{7a}
\includegraphics[width=6cm]{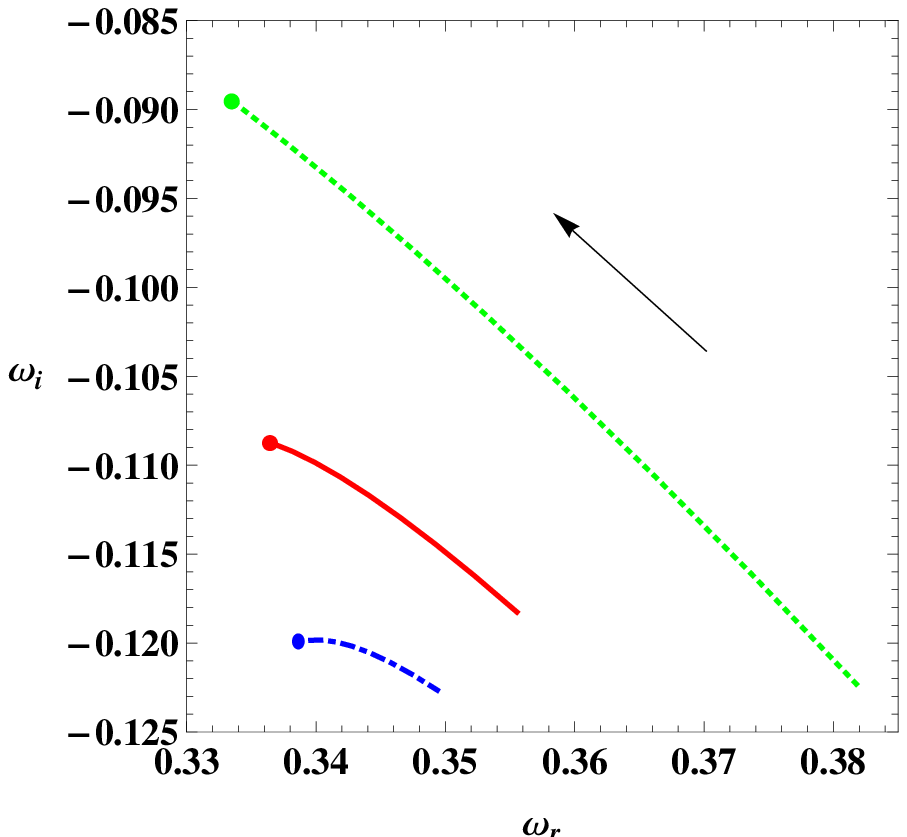}}
\subfigure[]{\label{7b}
\includegraphics[width=6cm]{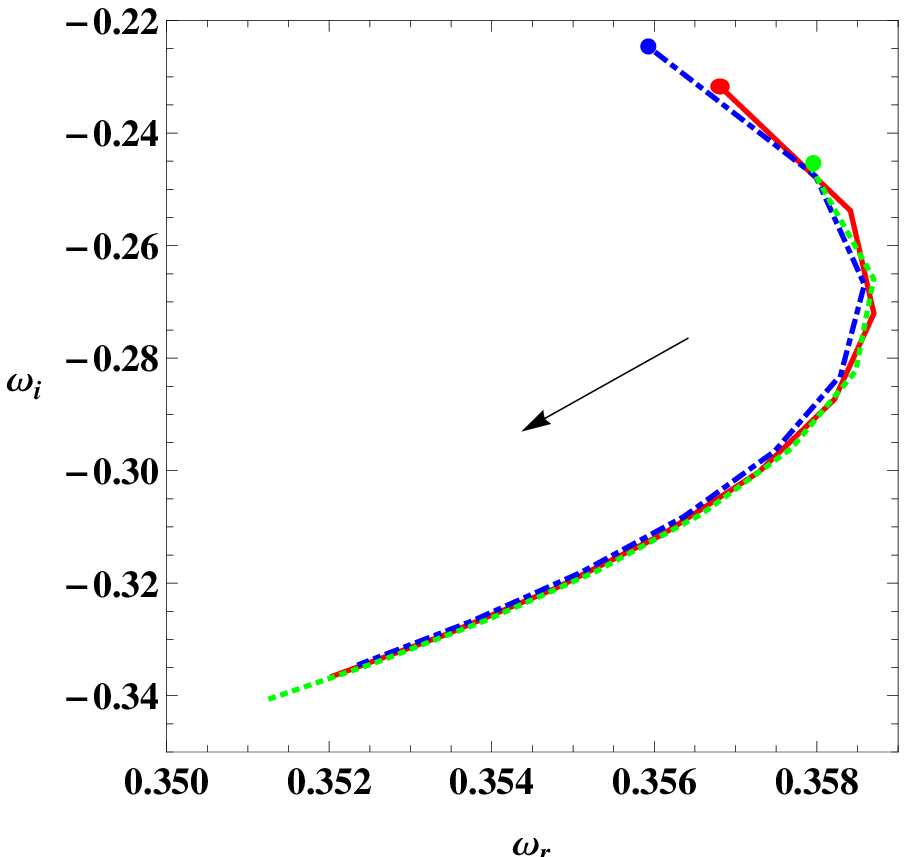}}
\caption{The behavior of the QNM frequencies in the complex-$\omega$ diagram with $T=0.04$, which is still below the critical temperature for $\lambda$=1.8, 2.0, and 2.2. (a) Small black holes. (b) Large black holes. The three lines correspond to $\lambda=1.8$ (green dotted line), $\lambda=2.0$ (red solid line) and $\lambda=2.2$ (blue dot dashed line). Solid point means phase transition point. The arrows indicate the increase of the black hole horizon.}\label{fig7}
\end{figure}

The behaviors of the QNM frequencies are given in Fig.~\ref{fig6} and Fig.~\ref{fig7}, for $T=0.03$ and 0.04, respectively. For low temperature, i.e., $T=0.03$, we can see that for the small black hole phase, the QNM frequencies have a negative slope, while for the large black hole phase, the slope is positive. Similar to the isobaric process, this also indicates that there exists a small/large black hole phase transition. Thus this method of sign change of slope can be used to probe the phase transition for low fixed temperature. On the other hand, at high fixed temperature, i.e., $T=0.04$, the case becomes different. The slope for the large black hole phase has an inflection point in the complex-$\omega$ diagram, see Fig.~\ref{7b}. Thus the slope method will be invalid for this case.

At the critical point $T=T_c$ with the values given in Table I, we find that the QNM frequencies of the two black hole phases possess the same behavior with the increase of the black hole horizon, which is just same as that shown in Fig.~\ref{fig4}. So for the isothermal process, this method of sign change of slope is also only valid at low fixed temperature compared with the critical temperature.

\section{Oscillatory behaviors of quasinormal modes}
\label{4}

As shown above, the method of sign change of slope on probing the black hole phase transition is only accurate for low fixed pressure or temperature case. While for the case of higher pressure or temperature very near the critical point, the method is problematic. So we would like to explore the detailed behavior of the QNM frequencies for the high fixed pressure or temperature case.

\begin{figure}
\subfigure[]{\label{8a}
\includegraphics[width=6cm]{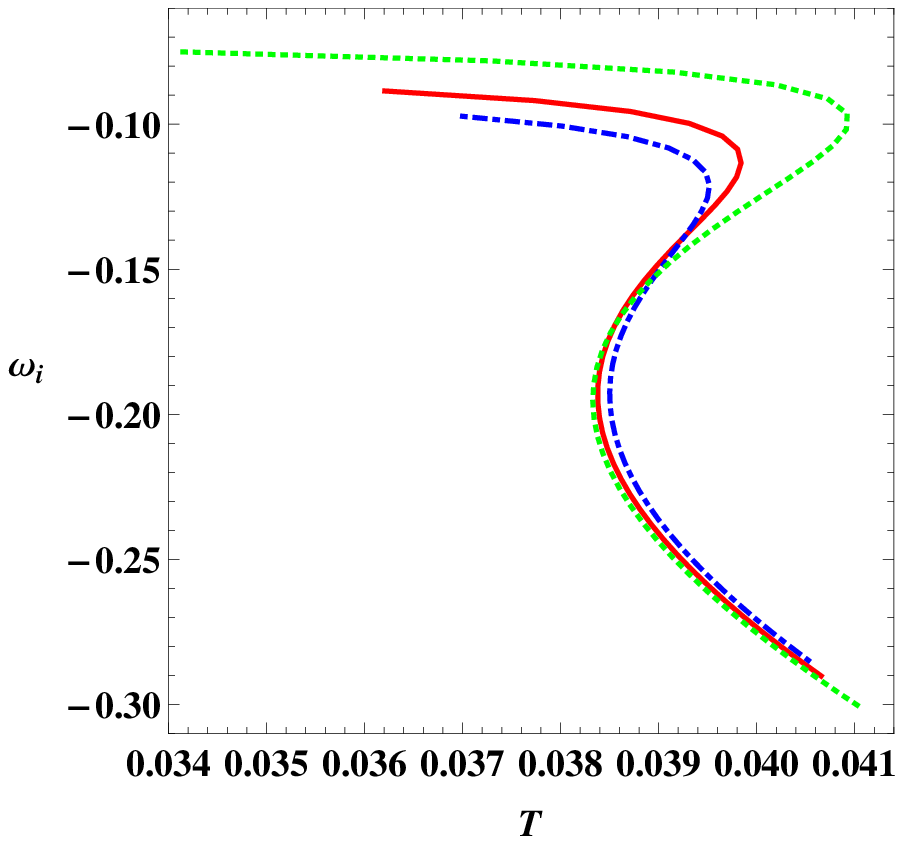}}
\subfigure[]{\label{8b}
\includegraphics[width=6cm]{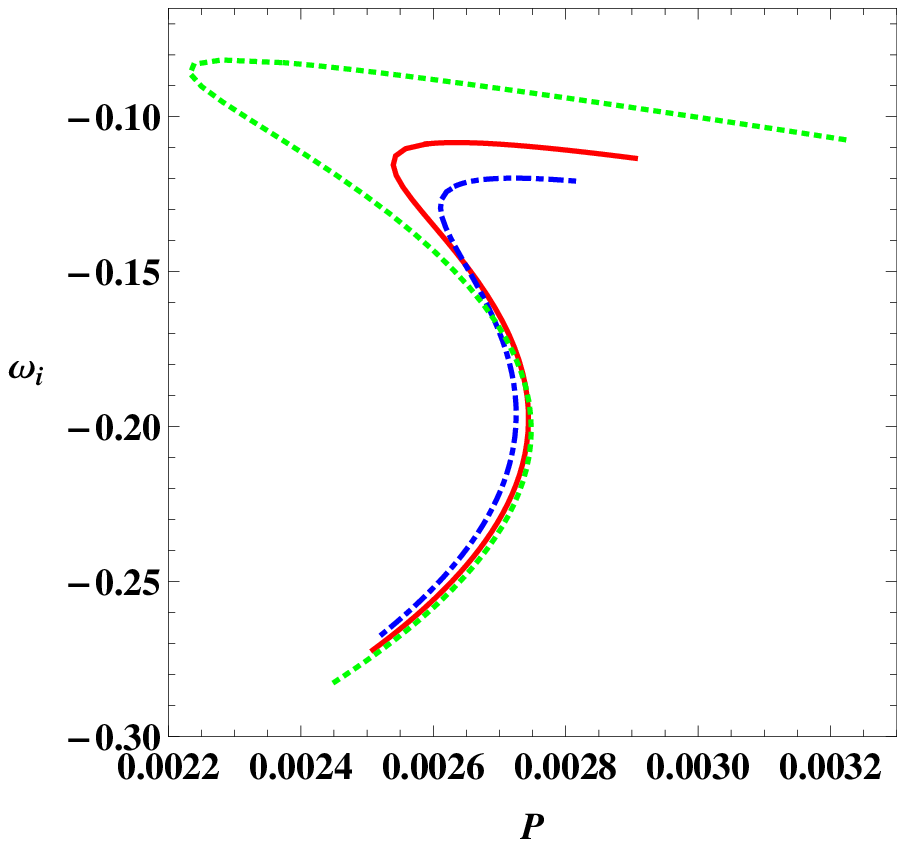}}
\caption{The non-monotonic behaviors of the QNM frequencies. (a) $\omega_i-T$ diagram with $P=0.0025$. (b) $\omega_i-P$ diagram with $T=0.04$. The three lines correspond to $\lambda=1.8$ (green dotted line), $\lambda=2.0$ (red solid line) and $\lambda=2.2$ (blue dot dashed line). The pressure and temperature we choose are below their critical values.}\label{fig8}
\end{figure}

\begin{figure}
\subfigure[]{\label{9a}
\includegraphics[width=5cm]{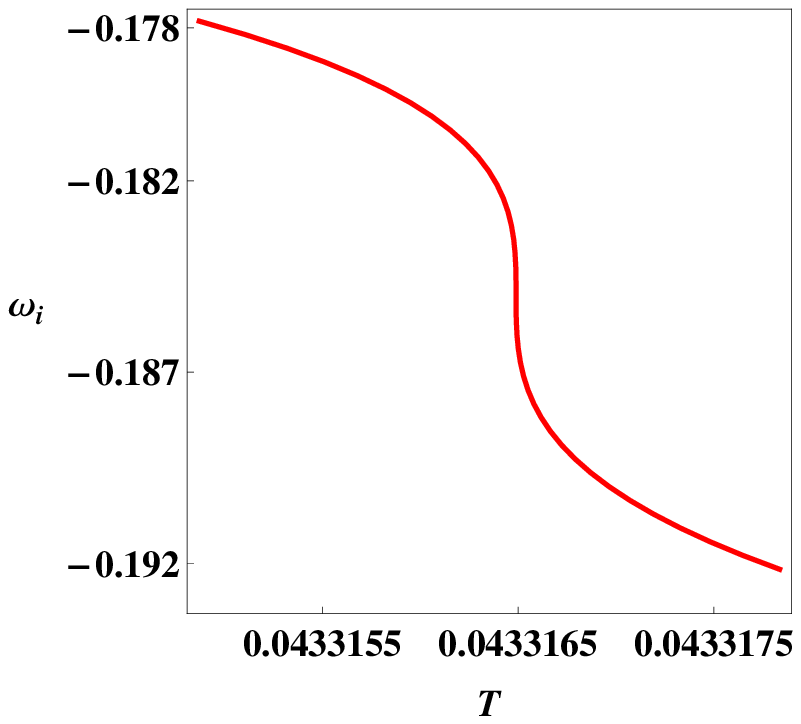}}
\subfigure[]{\label{9b}
\includegraphics[width=5cm]{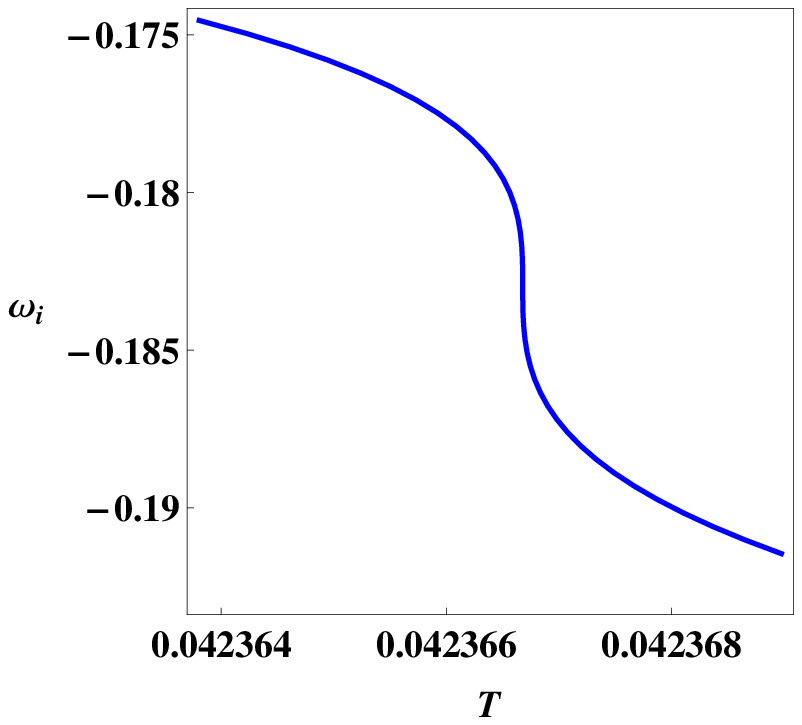}}
\subfigure[]{\label{9c}
\includegraphics[width=5cm]{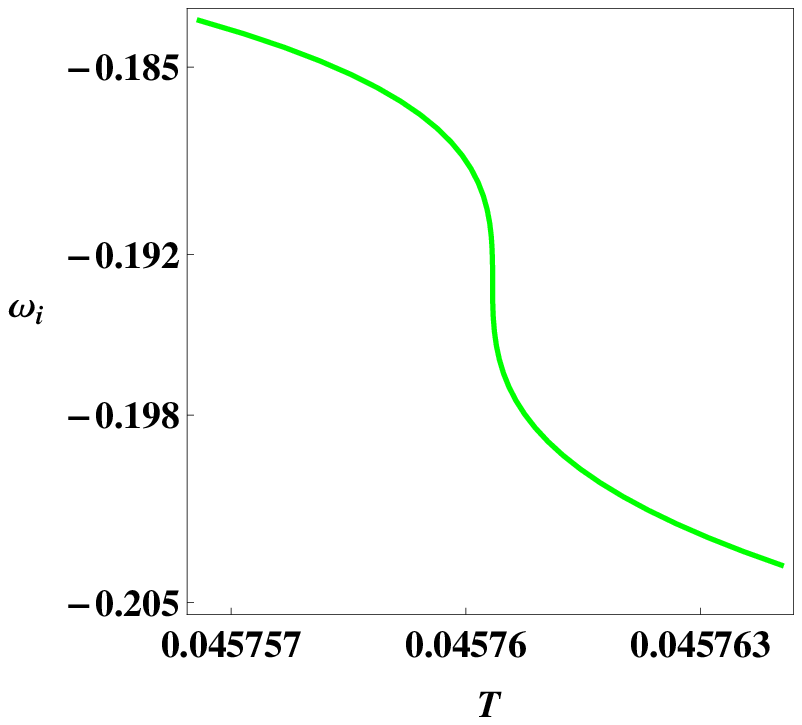}}\\
\subfigure[]{\label{9d}
\includegraphics[width=5cm]{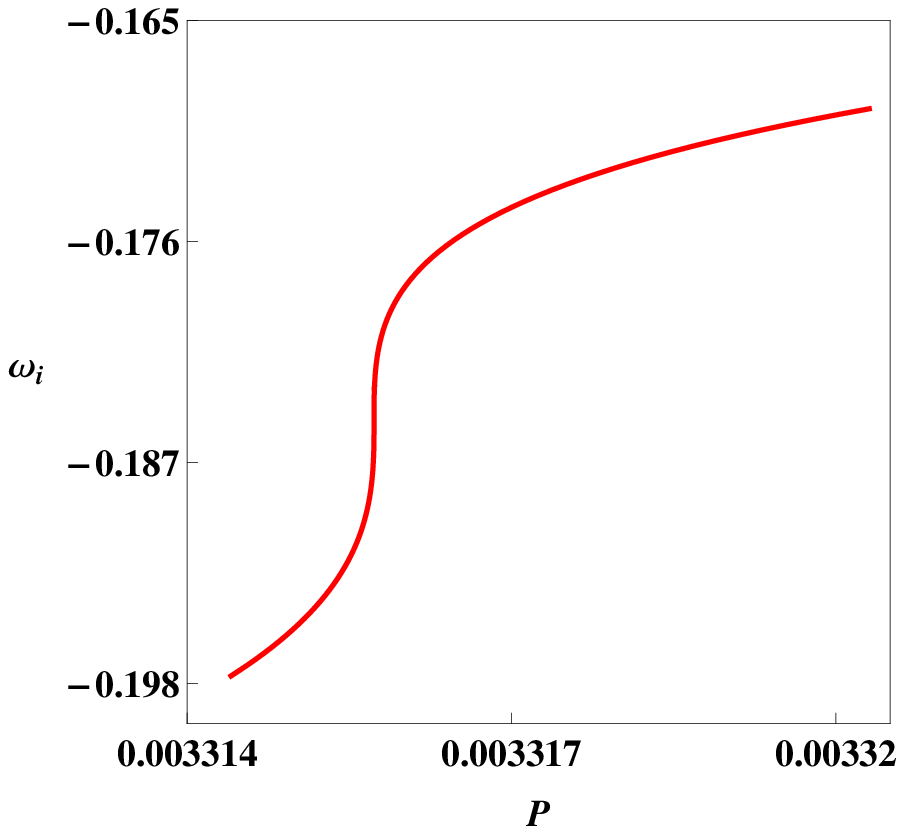}}
\subfigure[]{\label{9e}
\includegraphics[width=5cm]{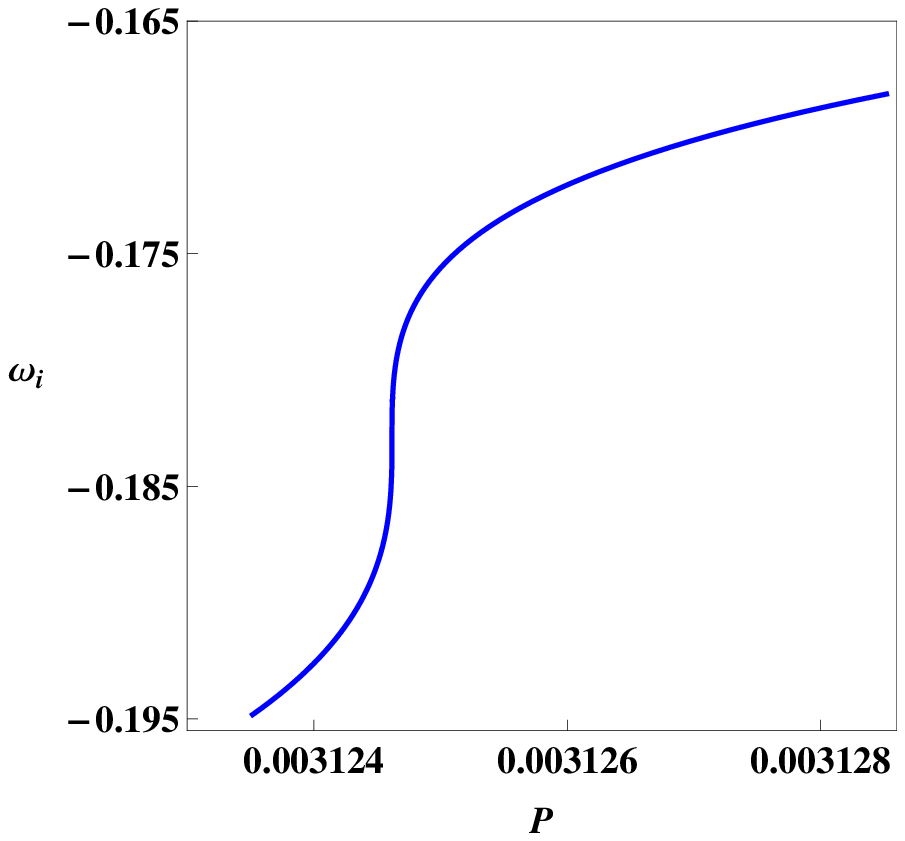}}
\subfigure[]{\label{9f}
\includegraphics[width=5cm]{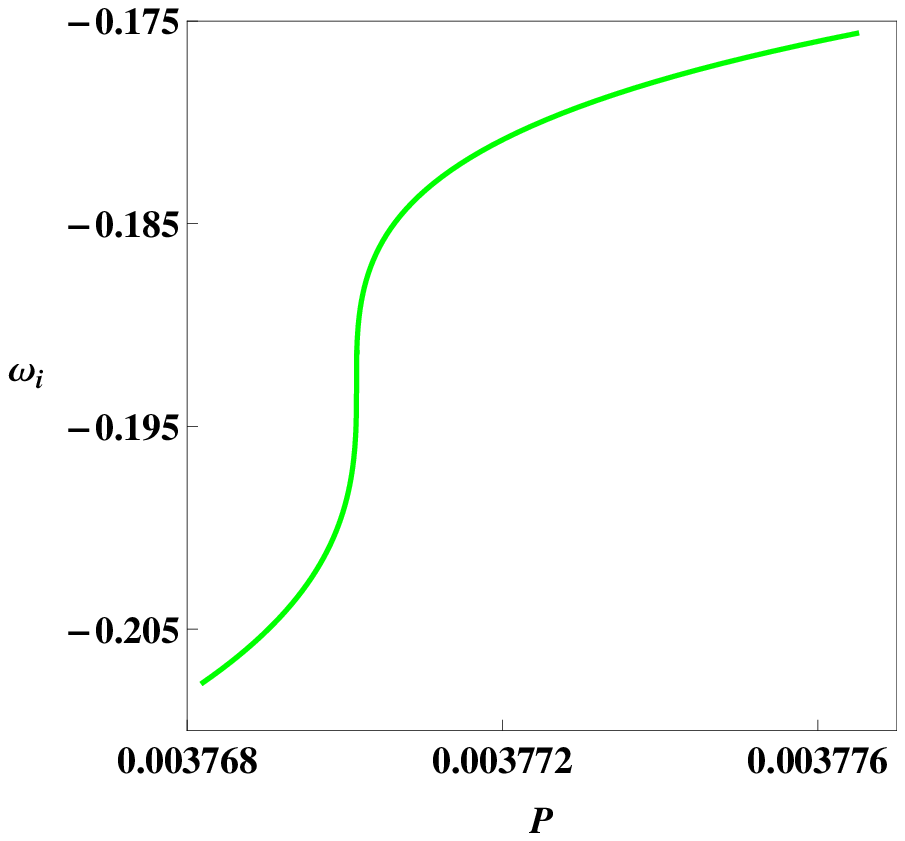}}
\caption{The non-monotonic behaviors of the QNM frequencies in the $\omega_i-T$ diagram (upper three panels) and $\omega_i-P$ diagram (lower three panels) at the critical point. The lines correspond to $\lambda=1.8$ (green dotted line), $\lambda=2.0$ (red solid line) and $\lambda=2.2$ (blue dot dashed line). The critical values can be found in Table I.}\label{fig9}
\end{figure}

\begin{figure}
\subfigure[]{\label{10a}
\includegraphics[width=6cm]{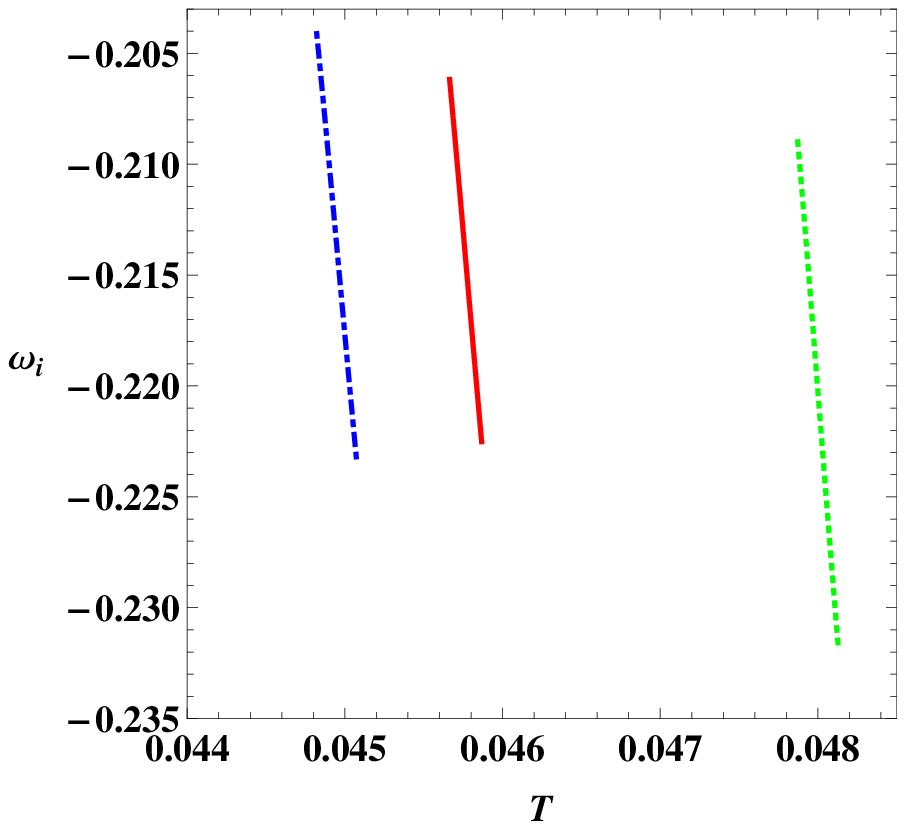}}
\subfigure[]{\label{10b}
\includegraphics[width=6cm]{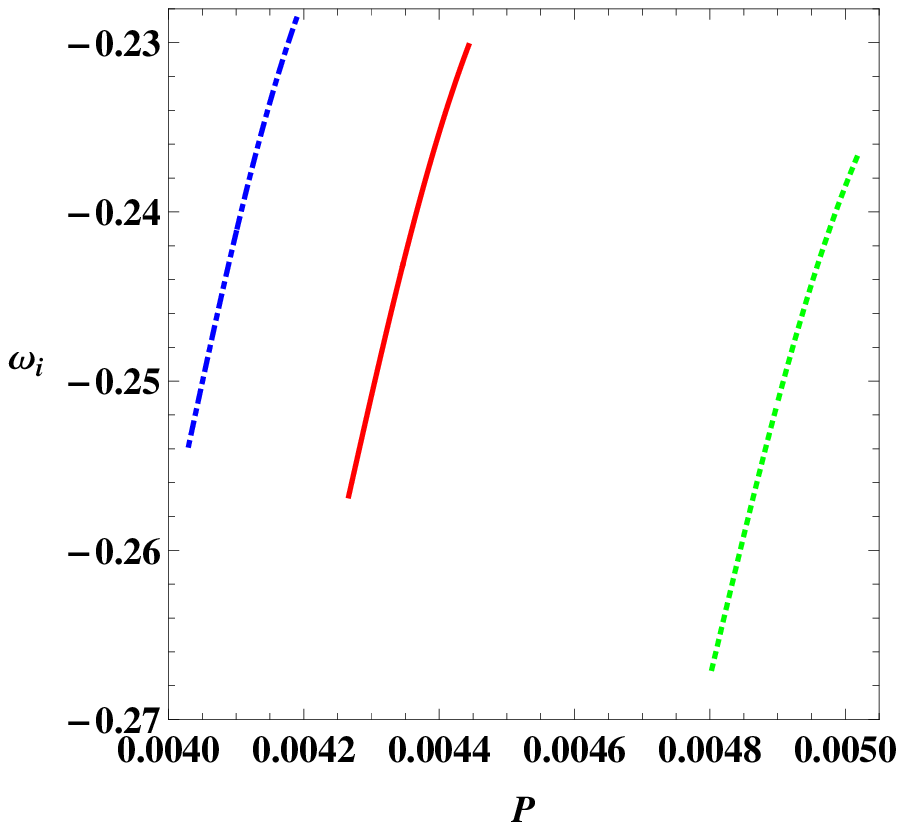}}
\caption{The behaviors of the QNM frequencies above the critical point. (a) $\omega_i-T$ diagram. (b) $\omega_i-P$ diagram. The three lines correspond to $\lambda=1.8$ (green dotted line), $\lambda=2.0$ (red solid line) and $\lambda=2.2$ (blue dot dashed line). The critical values can be found in Table I.}\label{fig10}
\end{figure}

Here we respectively plot the imaginary part of the QNM frequencies as a function of temperature and pressure in Fig.~\ref{fig8}. For each value of the parameter $\lambda$, there displays a non-monotonic behavior, which is similar to that of the vdW fluid. At the critical point, we show the behavior of $\omega_i$ in Fig.~\ref{fig9}. One can find that the non-monotonic behavior just disappears for different $\lambda$. For the isobaric processes, $\omega_i$ decreases monotonically with the temperature $T$. For the isothermal process, it increases monotonically with the pressure $P$. Beyond the critical point, the behavior is presented in Fig.~\ref{fig10}. It is clear that the non-monotonic behavior completely disappears. Considering these cases, we suggest that the non-monotonic behavior of $\omega_i$ can reveal the first-order thermodynamic phase transition of black holes in LSB massive gravity. Therefore, this non-monotonic behavior of $\omega_i$ is an additional probe for the black hole phase transition.

\section{Conclusions and Discussions}
\label{5}

In this paper, we calculated the QNM frequencies of the massless scalar perturbations around the black hole in the LSB massive gravity, and examined the relation between the QNM frequencies and thermodynamic phase transition. For $\gamma=-1$, the black hole system exhibits a small/large black hole phase transition of the vdW type. At low fixed temperature or pressure, we can see that there exists a slope change of the QNM frequencies for the small and large black hole phases both for the isobaric and isothermal processes. This result implies that we can use this method of sign change of slope to probe the black hole phase transition. However, when the temperature or pressure approaches certain value near the critical point, we found that there will be the inflection points in the complex-$\omega$ diagram. So, this method will be invalid. Moreover, at the critical point, the phase transition will be a second-order one. And for the small and large black holes, the behavior of the QNM frequencies will share the similar behavior, arguing that the method of sign change of slope is not appropriate to probe the second-order phase transition.

On the other hand, we also examined the behaviors of the imaginary part of the QNM frequencies along the isobaric and isothermal processes. For the fixed temperature or pressure below the critical value, the imaginary part clearly exhibits a non-monotonic behavior, see Fig.~\ref{fig8}, which is very similar to the vdW fluid. Thus, it may indicate the existence of the black hole phase transition. When increasing the temperature or pressure to its critical value, the non-monotonic behavior vanishes, see Fig.~\ref{fig9}. When further increasing the temperature or pressure, the non-monotonic behavior completely disappears. Therefore, the appearance of the non-monotonic behavior may imply the existence of the first-order phase transition. This also provides us with an additional method to probe the black hole thermodynamic phase transition through the QNMs.

\begin{table}[h]
\centering
\begin{tabular}{|c|c|c|c||c|c|c|c||c|c|c|c|}
\hline
\multicolumn{4}{|c|}{$\lambda=2.2$}&
\multicolumn{4}{|c|}{$\lambda=2.0$}&
\multicolumn{4}{|c|}{$\lambda=1.8$}\\ \hline
$T(10^{-2})$ & $r_h$ & $\omega_r$ & $\omega_{i}$
& $T(10^{-2})$ & $r_h$ & $\omega_r$ & $\omega_{i}$
& $T(10^{-2})$ & $r_h$ & $\omega_r$ & $\omega_{i}$ \\ \hline
2.38 & 1.299 & 0.26295&-0.04637 & 2.40 & 1.201 & 0.26265&-0.04268& 2.31& 1.046 &0.26385&-0.03649\\
2.64 & 1.349 & 0.26274&-0.04699 & 2.64 & 1.245 & 0.26248&-0.04320& 2.58& 1.083 &0.26369&-0.03689\\
2.85 & 1.399 & 0.26255&-0.04769 & 2.84 & 1.289 & 0.26232&-0.04379& 2.81& 1.121 &0.26353&-0.03735\\
3.01 & 1.449 & 0.26240&-0.04849 & 3.01 & 1.333 & 0.26218&-0.04446& 3.00& 1.158 &0.26340&-0.03787\\
3.14 & 1.499 & 0.26226&-0.04939 & 3.14 & 1.376 & 0.26206&-0.04522& 3.15& 1.196 &0.26327&-0.03845\\ \hline
3.14 & 6.441 & 0.30953&-0.37483 & 3.14 & 6.481 & 0.30930&-0.37489& 3.15& 6.607 &0.31069&-0.37631\\
3.29 & 7.441 & 0.32353&-0.39272 & 3.26 & 7.281 & 0.32097&-0.38902& 3.34& 7.807 &0.32945&-0.39957\\
3.46 & 8.441 & 0.34221&-0.41482 & 3.39 & 8.081 & 0.33496&-0.40601& 3.57& 9.007 &0.35322&-0.42776\\
3.67 & 9.441 & 0.36288&-0.43940 & 3.55 & 8.881 & 0.35098&-0.42491& 3.83& 10.207&0.38077&-0.45944\\
3.89 & 10.441& 0.38649&-0.46628 & 3.72 & 9.681 & 0.36877&-0.44568& 4.11& 11.407&0.41055&-0.49365\\ \hline
\end{tabular}
\caption{The QNM frequencies of the massless scalar perturbation with fixed pressure $P=0.0015$ below the critical pressures of $\lambda$=1.8, 2.0, and 2.2. The upper part, above the horizontal line, is for the small black hole phase, while the lower part is for the large black hole phase.}\label{tab1}
\end{table}

\begin{table}[h]
\centering
\begin{tabular}{|c|c|c|c||c|c|c|c||c|c|c|c|}
\hline
\multicolumn{4}{|c|}{$\lambda=2.2$}&
\multicolumn{4}{|c|}{$\lambda=2.0$}&
\multicolumn{4}{|c|}{$\lambda=1.8$}\\  \hline
$P(10^{-3})$ & $r_h$ & $\omega_r$ & $\omega_{i}$ &
$P(10^{-3})$& $r_h$ & $\omega_r$ & $\omega_{i}$ &
$P(10^{-3})$& $r_h$ & $\omega_r$ & $\omega_{i}$ \\ \hline
2.24 & 1.385 & 0.31094&-0.07877 & 2.48 &1.267& 0.32379&-0.08272& 3.15 &1.092& 0.35761&-0.09624\\
1.98 & 1.404 & 0.29495&-0.06799 & 2.15 &1.286& 0.30464&-0.06960& 2.61 &1.111& 0.33061&-0.07690\\
1.74 & 1.423 & 0.27961&-0.05837 & 1.85 &1.305& 0.28612&-0.05794& 2.13 &1.130& 0.30408&-0.05988\\
1.53 & 1.442 & 0.26489&-0.04980 & 1.58 &1.324& 0.26817&-0.04763& 1.71 &1.148& 0.27773&-0.04504\\
1.35 & 1.460 & 0.25073&-0.04221 & 1.35 &1.342& 0.25068&-0.03857& 1.34 &1.167& 0.25121&-0.03229 \\ \hline
1.35 & 6.979 & 0.26815&-0.21002 & 1.35 &7.014& 0.26821&-0.21126& 1.34 &7.170& 0.26817&-0.21447\\
1.30 & 7.479 & 0.26788&-0.21826 & 1.30 &7.514& 0.26790&-0.21935& 1.29 &7.670& 0.26778&-0.22217 \\
1.26 & 7.979 & 0.26741&-0.22551 & 1.26 &8.014& 0.26740&-0.22646& 1.25 &8.170& 0.26723&-0.22896 \\
1.22 & 8.479 & 0.26682&-0.23192 & 1.22 &8.514& 0.26679&-0.23277& 1.21 &8.670& 0.26658&-0.23500 \\
1.18 & 8.979 & 0.26615&-0.23763 & 1.18 &9.014& 0.26611&-0.23839& 1.17 &9.170& 0.26588&-0.24039 \\ \hline
\end{tabular}
\caption{The QNM frequencies of the massless scalar perturbation with fixed temperature $T=0.03$ below the critical temperatures of $\lambda$=1.8, 2.0, and 2.2. The upper part, above the horizontal line, is for the small black hole phase, while the lower part is for the large black hole phase.}\label{tab2}
\end{table}

\section*{Acknowledgements}

We thank Dr. Kai Lin for useful discussion. This work was supported by the National Natural Science Foundation of China (Grants No. 11675064, No. 11522541, No. 11375075, and  No 11205074), and the Fundamental Research Funds for the Central Universities (Grants Nos. lzujbky-2016-115).

\end{document}